\documentclass[aps,groupedaddress,twocolumn]{revtex4}
\usepackage{epsf}
\usepackage{epsfig}
\usepackage{graphicx}
\usepackage{color}
\usepackage{siunitx}
\usepackage{epstopdf}
\begin{document}

\title {
\large
Non-Sequential Double Ionization by Counter Rotating Circularly Polarized Two-Color Laser Fields}
\author{S. Eckart$^{1}$}
\author{M. Richter$^{1}$}
\author{M. Kunitski$^{1}$}
\author{A. Hartung$^{1}$}
\author{J. Rist$^{1}$}
\author{K. Henrichs$^{1}$}
\author{N. Schlott$^{1}$}
\author{H. Kang$^{1,2}$}
\author{T. Bauer$^{1}$}
\author{H. Sann$^{1}$}
\author{L.Ph.H. Schmidt$^{1}$}
\author{M. Sch\"offler$^{1}$}
\author{T. Jahnke$^{1}$}
\author{R.~D\"{o}rner$^{1}$}
\email{doerner@atom.uni-frankfurt.de}

\affiliation{
$^1$ Institut f\"ur Kernphysik, Goethe-Universit\"at, Max-von-Laue-Str. 1, 60438 Frankfurt, Germany \\
$^2$ State Key Laboratory of Magnetic Resonance and Atomic and Molecular Physics, Wuhan Institute of Physics and Mathematics, Chinese Academy of Sciences, Wuhan 430071, China \\
}

\date{\today}

\begin{abstract}
We report on non-sequential double ionization of Ar by a laser pulse consisting of two counter rotating circularly polarized fields ($\SI{390}{\nano\meter}$ and $\SI{780}{\nano\meter}$). The double ionization probability depends strongly on the relative intensity of the two fields and shows a ``knee''-like structure as function of intensity. We conclude that double ionization is driven by a beam of nearly monoenergetic recolliding electrons, which can be controlled in intensity and energy by the field parameters. The electron momentum distributions show the recolliding electron as well as a second electron which escapes from an intermediate excited state of Ar$^+$.

\end{abstract}
\maketitle

\begin{figure*}
\includegraphics[width=17.9cm]{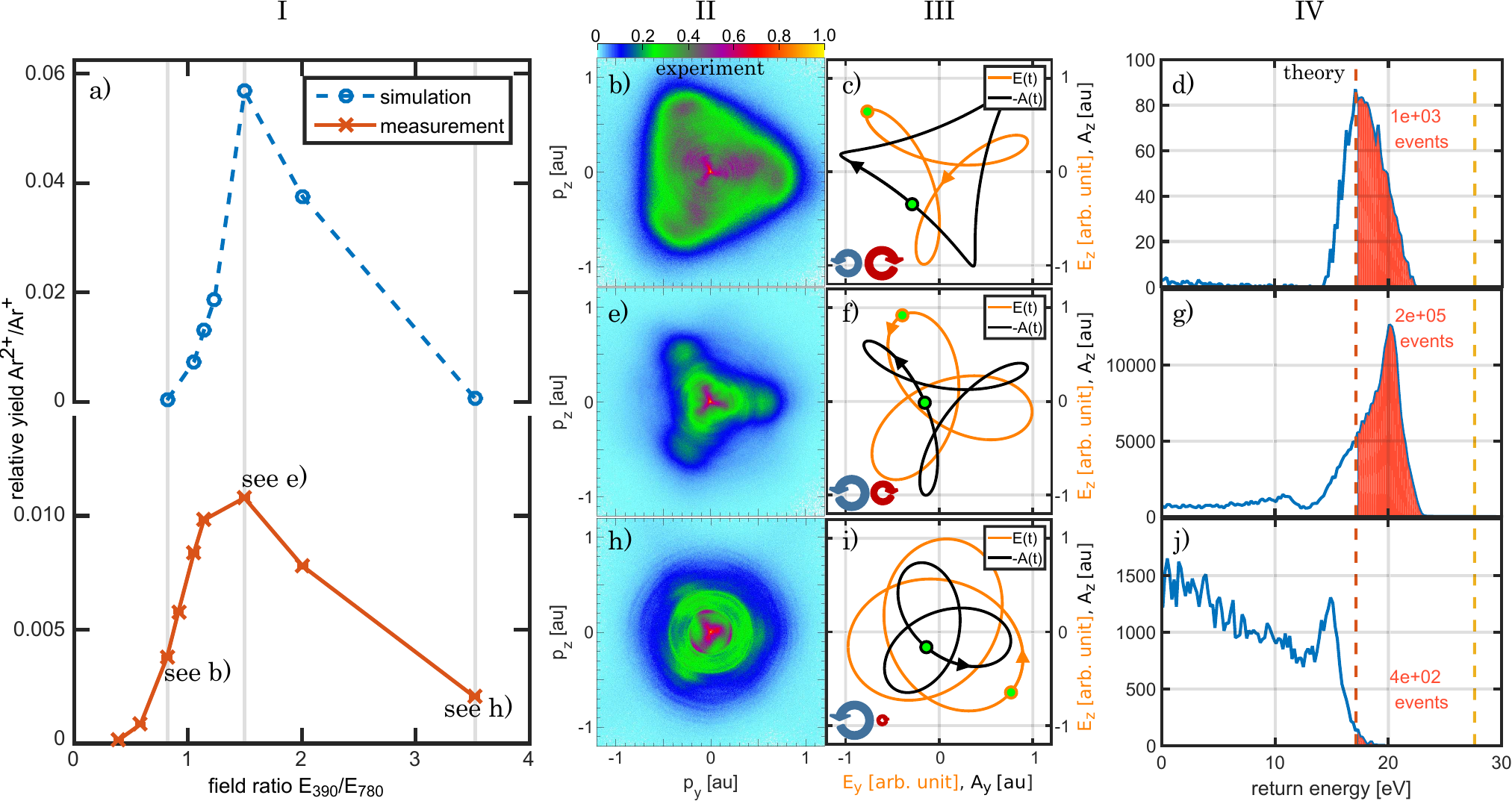}
\caption{The measured and simulated relative yield of double to single ionization is shown in column I. In column II measured electron momentum distributions are shown for the field ratios $E_{390}/E_{780}$ of (b) 0.8:1, (e) 1.5:1 and (h) 3.5:1 and a maximum combined intensity of $\SI{5.0E+14}{\watt\per\centi\meter^2}$. In column III the corresponding combined electric fields $E_{y/z}$ (arb. units) as well as the vector potentials $A_{y/z}$ (atomic units) are depicted for the same ratios $E_{390}/E_{780}$ (c) 0.8:1, (f) 1.5:1 and (i) 3.5:1. The phase in which the combined electric field is at one of its maxima is marked with a green dot. The helicities of the two colors and the temporal development of the combined electric fields and the vector potentials are indicated with arrows.  In IV the corresponding simulated energy distribution of the returning electrons is shown which has been calculated also for the ratios $E_{390}/E_{780}$ (d) 0.8:1, (g) 1.5:1 and (j) 3.5:1 using CTMC. The energy needed for double ionization ($\SI{27.6}{\electronvolt}$) and the relevant excitation energy ($\SI{17.14}{\electronvolt}$) are marked with a dashed orange and a red line, respectively. The number of recollision events is normalized to $10^6$ single ionization events in IV. All events exceeding $\SI{17.14}{\electronvolt}$ are integrated which is visualized as the red shaded area. This integral corresponds to the number of returning electrons having enough energy to excite $\text{Ar}^+$ and strongly depends on the field ratio $E_{390}/E_{780}$.}
\label{fig_figure1}
\end{figure*}

Strong laser fields efficiently lead to the ejection of electrons from atoms and molecules. In the continuum the electron wave packet is driven by the laser field and its trajectory can be controlled by tailoring the time evolution of the electric field vector of the laser pulse on a sub cycle basis. A laser pulse composed of two harmonic colors offers already a significant amount of control parameters, such as polarization, relative intensity and phase between the two fields. This allows shaping the light field and thus to steer the electron motion in the continuum or in a bond \cite{Kitzler2005prl,Zhang2014prl,Richter2015,Geng2015,Douguet2016,Mancuso2015,Medisauskas2015,Hasovic2016,lin2016jpb}. Particularly versatile and in addition well controllable waveforms are generated by counter rotating circular two-color fields (CRTC) shown in Fig. 1 panels (c), (f), (i). These waveforms have spawned recent activities because, unlike circularly polarized light consisting of only a single color, CRTC fields can initially drive electrons away from the atom they have escaped from but later drive them back - often on triangularly shaped trajectories to re-encounter their parent ion. The recapture of these electrons gives rise to the emission of circularly polarized higher harmonic light as predicted in pioneering work by Becker and coworkers \cite{becker1999schemes} and confirmed by recent experimental studies \cite{fleischer2014spin}. The recollision in such fields has also been identified by high energetic electrons \cite{mancuso2016controlling} as well as by characteristic structures in the electron momentum distribution at very low energies where the electrons are Coulomb focused \cite{Coulomb_focusing}.

In the present work we experimentally show that CRTC fields also lead to efficient double ionization mediated by the re-colliding electron as predicted by recent classical ensemble calculations \cite{chaloupka2016}. Studying the probability of double ionization and in particular the three dimensional momentum distribution of the emitted electrons gives unprecedented insight into the recollision dynamics occurring in these two-color laser fields. In particular they support that CRTC fields can be used to create a nearly monoenergetic electron beam for attosecond time-resolved studies. Additionally, very recent theoretical work \cite{milovsevic2016possibility} building on \cite{olga_spinpolarized} predicts that these recolliding electrons can be generated such that they are spin polarized \cite{alex_in_press}.

\begin{figure}[ht]
\includegraphics[width=7.8cm]{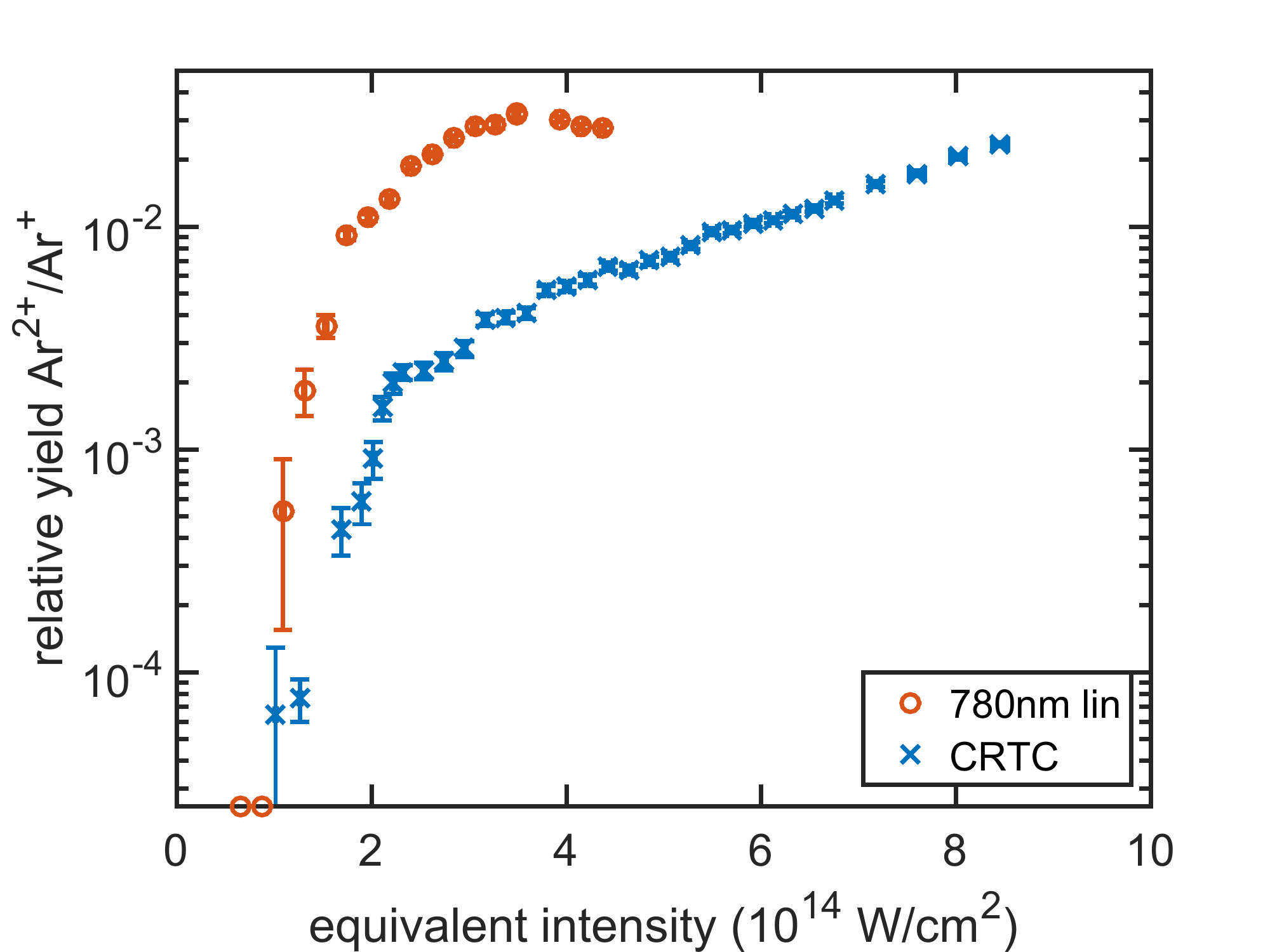}
\caption{Relative yield of double ionization is shown as a function of intensity in focus for linear polarization at $\SI{780}{\nano\meter}$ and for a CRTC field with $E_{390}/E_{780}=1.23$ which is close to the maximum in Fig. \ref{fig_figure1} (a). For the CRTC field a similar ``knee''-like shape is observed as it is commonly known for linear polarization. The two first points for linear polarization on the ordinate indicate that no double ionization events were measured within a reasonable amount of time.}
\label{fig_intensity_scan}
\end{figure}

In order to generate two-color fields, we use a $200$ $\mu\text{m}$ BBO to double the frequency of a $\SI{780}{\nano\meter}$ laser pulse (KMLabs Dragon, $\SI{40}{\femto\second}$ FWHM, $\SI{8}{\kilo\hertz}$). The fundamental and the second harmonic are separated using a dielectric beamsplitter. Before the two are recombined a neutral density filter, followed by a lambda-half and a lambda-quarter waveplate are installed in each pathway. In the arm of the fundamental wavelength a nm-delay stage (PI P-752.2CD) is used to adjust the temporal overlap and the relative phase between the two colors. Scanning the relative phase between the fundamental and the second harmonic actively using the nm-delay stage allows to correct for slow phase drifts in the offline analysis \cite{Richter2015}. Using a spherical mirror ($f=\SI{80}{\milli\meter}$) inside the vacuum chamber we reach intensities of up to $\SI{1.9E+14}{\watt\per\centi\meter^2}$ for the fundamental and $\SI{2.6E+14}{\watt\per\centi\meter^2}$ for the second harmonic.
The laser field is focused into an argon target that is generated using supersonic gas expansion. The peak intensity in the focus was calibrated separately for both colors using the photoelectron momentum distributions from ionization by circularly polarized light. For $\SI{780}{\nano\meter}$ the intensity in the focus was obtained from the measured drift momentum. For $\SI{390}{\nano\meter}$ we observed clear peaks in the photoelectron energy distribution spaced by the photon energy. We used the shift of the energy of these peaks as function of intensity, which is due to the change in ponderomotive energy, for calibration of the intensity in the focus. The uncertainty of the absolute intensity for $\SI{780}{\nano\meter}$ and $\SI{390}{\nano\meter}$ is estimated to be $10\,\%$ and $20\,\%$ respectively. Further it should be noted that our intensity calibration is an insitu measurement of the electric field and hence includes focal averaging for single ionization. To minimize the effect of focal averaging the gas jet was collimated to $40\pm 10$ $\mu\text{m}$ along the axial direction in the laser focus (Rayleigh length of about $\SI{150}{\micro\meter}$) for all measurements except the data presented in Fig. \ref{fig_intensity_scan} (in this case double ionization rates would have been too low for low intensities).

The three-dimensional electron momenta presented in this paper have been measured in coincidence with argon ions using cold-target recoil-ion momentum spectroscopy (COLTRIMS) \cite{ullrich2003recoil}. The length of the electron and ion arm was $\SI{378}{\milli\meter}$ and $\SI{67.8}{\milli\meter}$ respectively. The homogeneous electric and magnetic fields of $\SI{11.2}{\volt\per\centi\meter}$ and $8.3\,$G, respectively, guided electrons and ions towards position sensitive microchannel plate detectors with three layer delay-line anodes \cite{jagutzki2002multiple}. Only one electron was measured for single and double ionization events.

Using momentum conservation as criterion the false electron-ion coincidences for single ionization of argon resulting from simultaneous ionization of two different atoms in one laser shot are determined to be $25\%$. This leads to $14\%$ of the electrons that are detected in coincidence with $\text{Ar}^{2+}$ to stem from simultaneous single ionization of a second atom. We have corrected our measured electron momentum distributions for double ionization for these false coincidences.

In Fig. \ref{fig_figure1} (a) we show the ratio of doubly to singly charged argon ions as function of the relative field strength $E_{390}/E_{780}$ of the $\SI{780}{\nano\meter}$ and $\SI{390}{\nano\meter}$ field. The corresponding photoelectron distributions (b), (e), (h) are presented in Fig. \ref{fig_figure1} along with combined electric fields and vector potentials (c), (f), (i) for three selected field ratios. The maximum combined electric field amplitude corresponds to an intensity of $\SI{5.0E+14}{\watt\per\centi\meter^2}$. To characterize the strength of the different combined fields we use ``equivalent intensities'', which ensures that the single ionization rate for given ``equivalent intensities'' is similar for different shapes of the field. We find that the double ionization probability depends strongly on the relative intensity of the two-colors confirming a recent prediction \cite{chaloupka2016}. The double ionization probability shows a maximum at the ratio $E_{390}/E_{780}=1.5$. Double ionization vanishes for purely circular light which is a clear evidence that double ionization takes place non-sequentially and is mediated by a recolliding electron. 

To illuminate the recollision physics causing the observed dependence of double ionization probability we show the measured electron momentum distributions for single ionization [Fig. \ref{fig_figure1}, column II] together with the simulated energy distributions of recolliding electrons [Fig. \ref{fig_figure1}, column IV]. The measured electron momentum distributions exhibit a strong dependence on $E_{390}/E_{780}$ in qualitative agreement with recent reports by Mancuso et al. \cite{Mancuso2015,mancuso2016controlling}. The discussion of single ionization is not in the focus of the present paper but is shown here to guide intuition. Panels (d), (g), (j) in Fig. \ref{fig_figure1} show simulated energy distributions of recolliding electrons at the distance of closest approach using a classical trajectory Monte Carlo (CTMC) simulation. For this we have calculated classical electron trajectories with weights obtained by ADK theory \cite{Delone1991} starting at the exit of the tunnel with zero momentum in direction parallel to the tunnel and a Gaussian distribution of momenta transverse to the tunnel direction as given by ADK theory. The electron trajectories are calculated in the presence of the Coulomb field. In order to select trajectories which potentially can lead to ionization or excitation of the parent ion by energy transfer from the recolliding electron, we have postselected electrons, which have passed the parent ion at a distance smaller than $d_{min}=5\,$a.u. and have plotted their excess energy at the instant of closest approach. The number of trajectories has been normalized to the total number of $10^6$ freed electrons and thus giving the fraction compared to total single ionization. Also indicated in Fig. \ref{fig_figure1} (column IV) are the energy of the first relevant excited state of $\text{Ar}^+$ at $\SI{17.14}{\electronvolt}$ and the ionization threshold at $\SI{27.6}{\electronvolt}$. Strikingly, the distributions of recollision energies shows a pronounced narrow peak which is swept in energy and intensity as the ratio $E_{390}/E_{780}$ is varied. The measured maximum of the double ionization yield [Fig. \ref{fig_figure1} (a)] coincides with the field ratio for which the flux of recolliding electrons with energies sufficient to excite the $\text{Ar}^+$ reaches its maximum. At the intensities used in the experiment there are no recolliding electrons with energies above the $\text{Ar}^+$ ionization threshold (yellow line in Fig. \ref{fig_figure1} column IV). This suggests that double ionization occurs solely by recollision induced excitation followed by subsequent ionization of $\text{Ar}^{+*}$ (see e.g. \cite{weber2000correlated,Feuerstein2001prl,liu2008strong}). 

To model the double ionization yield by impact excitation more quantitatively we have weighted the simulated electron energy distributions shown in Fig. \ref{fig_figure1} (d), (g), (j) with the energy dependent electron impact excitation cross section from \cite{Jesus2004} and integrated over the electron energy. This simple estimate reproduces the observed $E_{390}/E_{780}$ dependence of double ionization as shown by the dashed line in Fig. \ref{fig_figure1} (a). Although the obtained maximum turned out to be robust regarding the choice of $d_{min}$ it should be noted that the somewhat arbitrary value of $d_{min}$ as selection criterion of the recollision trajectory allows us to estimate only the dependence of the double ionization yield on $E_{390}/E_{780}$ but not an absolute value as for example given by quantitative rescattering (QRS) theory \cite{chen2009quantitative}.

So far our scan of $E_{390}/E_{780}$ together with the trajectory simulations suggests that double ionization in CRTC fields is driven by recollision induced excitation and that tuning $E_{390}/E_{780}$ allows to create nearly monoenergetic beams of recolliding electrons [see Fig. \ref{fig_figure1} (d), (g)]. To further support and detail these findings we discuss in the remainder of this paper the intensity dependence of double ionization in CRTC fields and compare the three dimensional momentum distribution of the electrons from single and double ionization.

Fig. \ref{fig_intensity_scan} shows the intensity dependence of the $\text{Ar}^{2+}/\text{Ar}^{+}$ ratio for a CRTC field with $E_{390}/E_{780}=1.23$ which is close to the maximum in Fig \ref{fig_figure1} (a). We have found a steep increase of double ionization at an intensity of around $\SI{2.0E+14}{\watt\per\centi\meter^2}$ followed by a broad plateau. This shape is very similar to the ``knee''-like shape for linear polarized light (red circles) (see e.g. \cite{walker1994precision}). The double ionization probability of the CRTC fields is suppressed by one order of magnitude compared to linearly polarized light. The reason for this suppression is that in contrast to linearly polarized light the electrons in CRTC fields are driven in two dimensions and hence they often require some initial momentum in order to recollide with the core. This is similar to the double ionization suppression for elliptically polarized light or orthogonal linearly polarized two-color fields \cite{Zhang2014prl}. An inspection of the energy of simulated recollision electrons confirms, that the sharp drop-off of double ionization below $\SI{2.0E+14}{\watt\per\centi\meter^2}$ is caused by an absence of recolliding electrons with energies above the first relevant excitation energy of $\text{Ar}^+$ at $\SI{17.14}{\electronvolt}$. The rise of double ionization beyond $\SI{6.0E+14}{\watt\per\centi\meter^2}$ is caused by the onset of sequential ionization.

\begin{figure}[ht]
\includegraphics[width=8.5cm]{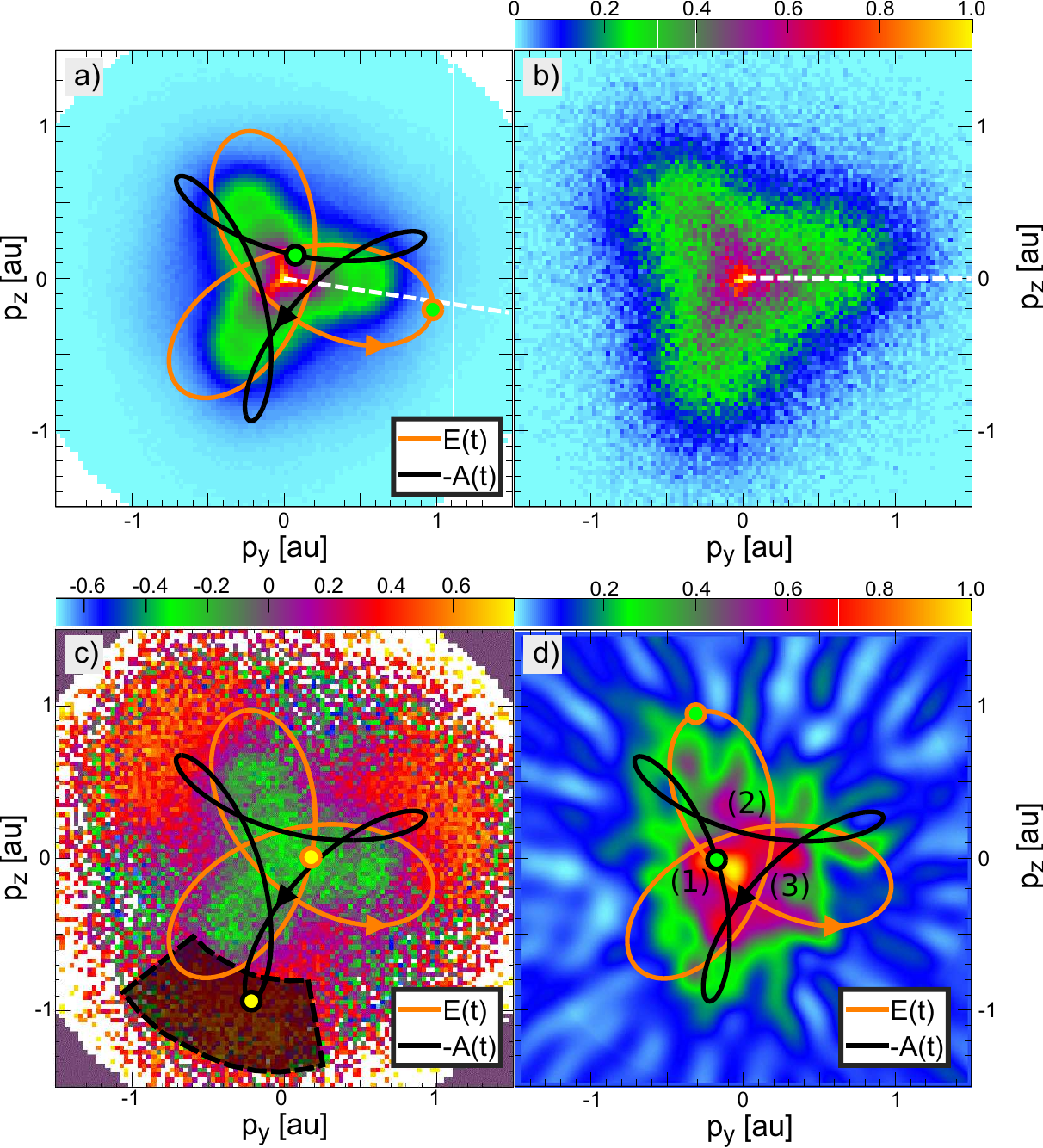}
\caption{Measured electron momentum distributions for single (a) and double (b) ionization for a ratio of $E_{390}/E_{780}=1.42$ and a combined intensity of $\SI{4.5E+14}{\watt\per\centi\meter^2}$. The corresponding vector potential is shown in black and the phase that belongs to the maximum combined field is indicated by a green dot. The dashed white line serves as guide to the eye regarding the rotation of the low energy structure (purple structure at momenta smaller than $0.3$ au). (c) shows the normalized difference $\frac{A-B}{A+B}$ where A (B) is the number of counts in the corresponding bins normalized to the sum of all counts for double (single) ionization. In (d) the tomographically reconstructed electron momentum distribution is shown for the condition that the first electron has been detected in the shaded area in (c). Electrons in this area have recollisonally excited the ion at the time of the pulse which is indicated by the yellow dot in c). d) shows the momenta of the corresponding excited electron set free later in the pulse. The labels 1, 2, 3 the first (1) second (2) or third (3) maximum of the field following the recollision and correspond to time delays of $433$, $1300$, $2167$ attoseconds after recollision.}
\label{fig_prl_figure3}
\end{figure}

To further illuminate the double ionization dynamics in the CRTC fields we have measured the electron momentum distribution for double ionization [Fig. \ref{fig_prl_figure3} (b)] and compare it to the one for single ionization [Fig. \ref{fig_prl_figure3} (a)]. First inspection shows significantly higher momenta and a counter-clockwise rotation of the triangular structure for electrons from double ionization as compared to those from single ionization. The counter-clockwise rotation is caused by the increase of the influence of the Coulomb potential for the doubly charged as compared to the singly charged ion. Single ionization electrons freed at the maximum of the field with zero energy would in the absence of the Coulomb potential be streaked to a certain final momentum. This final momentum is given by the negative vector potential at that time shown by the green dot on the black line. We have taken the rotation angle of the Lissajous curve relative to our measured electron momentum distribution from TDSE calculations \cite{mancuso2016controlling} and our own classical simulations. (Due to Coulomb effects this absolute rotation angle should not be taken as a quantitative measure.)

From our arguments given in the discussion of Fig. \ref{fig_figure1} we expect two classes of electrons for double ionization: the initial electron, which recollides and creates an excitation and the second electron, which is freed from the excited state later on in the laser field. The recolliding electron can be best seen in the data by calculating the normalized differences of the momentum distribution from double and single ionization [Fig.  \ref{fig_prl_figure3} (c)]. The yellow reddish area at the three tips of the vector potential in Fig. \ref{fig_prl_figure3} (c) results from these recolliding electrons. 
The feature spreads by $0.4\,$-$\, 0.5\,$a.u. around the vector potential, indicating an electron momentum after rescattering which is in agreement with an energy excess of $\SI{3}{\electronvolt}$ left on the electron after excitation. 

From the location of this feature close to the maximum of the vector potential [marked with a yellow dot in Fig. \ref{fig_prl_figure3}  (c)] we conclude that the main contribution results from recollisions close to the electric field minimum. The second electron, which is set free at a subsequent maximum of the electric field [corresponds to a minimum of the vector potential and is marked with a green dot in Fig. \ref{fig_prl_figure3}  (a), (d)], can be expected to have a momentum distribution similar to that of single ionization, with the main difference that it shows a bigger Coulomb effect escaping from a doubly charged ion. Also this is clearly seen in the data as the inner threefold structure in Fig. \ref{fig_prl_figure3} (b) is rotated counter-clockwise compared to the one from single ionization (a). The rotation is indicated by the dashed white line serving as guide to the eye (see e.g. \cite{spanner2012coulomb}). 

In the next step, we now investigate the time delay between the recollisional excitation and emission of the excited electron. To do so we impose the condition that the detected electron's momentum is in the gray shaded area in Fig. \ref{fig_prl_figure3} (c); we know that those electrons are mainly recolliding electrons. This sets the time zero for our time measurement. For technical reasons such as detector dead-time and poor ion momentum resolution in the two spatial dimensions we have used a filtered back-projection algorithm to tomographically reconstruct the momentum distribution \cite{smeenk2009momentum,Mancuso2015,mancuso2016controlling} of the second electron from the sum of the momenta of the first electron and the ion in z-direction as well as the relative phase of the two colors (determines the rotation of the combined electric field in space). High frequencies are cut off in frequency space. The result is shown in Fig. \ref{fig_prl_figure3} (d). It can be nicely seen that the second electron is ionized close to the first [marked with (1) in Fig. \ref{fig_prl_figure3} (d)]  maximum of the electric field that corresponds to a minimum of the vector potential, about $433$ attosconds after the recolliding electron has hit the ion. At the second field maximum (label (2)) the electron flux is already reduced and by the time of the third maximum after $2167$ attoseconds only a small fraction of excited electrons is left. This is a powerful example of how to use CRTC fields to study electron-emission dynamics on the attosecond timescale without being limited to near-single-cycle laser pulses (see e.g. \cite{bergues2012attosecond}).

In conclusion, we have shown that CRTC fields support non-sequential double ionization and at the same time allow for a detailed control of the energy distribution of the recolliding electron wave packet which re-encounters the parent ion from defined directions. These recolliding electron wave packets moving in CRTC fields have exciting applications. One is their use for the generation of higher harmonics with circular polarization, which is exploited already today \cite{fleischer2014spin}. Another possible application is the use of these wave packets to induce molecular dynamics by recollision \cite{Xie2012prl} where the CRTC field would allow to manipulate the energy and the direction from which the electron wave packet approaches the molecule. For such applications it is particularly useful that due to the shape of the vector potential the rescattered electrons are driven to an otherwise mainly empty region of momentum space as demonstrated by our measurement. Yet another exciting future application builds on the fact that at the time the electron is set free the field vector rotates as in circularly polarized light. This rotation gives rise to spin polarization of the emitted electrons as predicted in pioneering work \cite{olga_spinpolarized} and recently shown experimentally \cite{alex_in_press}. Our study shows that CRTC fields are highly efficient to generate such electrons for electron impact ionization or excitation paving the way for attosecond time-resolved collision studies with spin polarized electrons \cite{milovsevic2016possibility}. 
\small
\section{Acknowledgments}
\normalsize
This work was supported by the DFG Priority Programme ``Quantum Dynamics in Tailored Intense Fields''. We thank Xiao-Min Tong for providing results from his TDSE calculations in numerical form.
\bibliographystyle{unsrt}

\end{document}